\documentclass[aps, prl, floatfix, nofootinbib, twocolumn]{revtex4-2}

\usepackage{latexsym}
\usepackage{amsmath}
\usepackage{amssymb}
\usepackage{amsfonts}
\usepackage{textcomp}
\usepackage{color}
\usepackage{CJKutf8}

\usepackage[mathscr,scaled=1.15]{urwchancal}
\DeclareFontFamily{OT1}{pzc}{}
\DeclareFontShape{OT1}{pzc}{m}{it}%
{<-> s * [1.15] pzcmi7t}{}
\DeclareMathAlphabet{\mathpzc}{OT1}{pzc}{m}{it}

\usepackage{color}

\usepackage{supertabular}
\usepackage{placeins}
\usepackage{epsfig}
\usepackage{graphicx}

\definecolor{purple}{rgb}{0.5,0,0.5}
\definecolor{blue}{rgb}{0.0,0,0.9}
\definecolor{prdblue}{rgb}{0.133,0.118,0.498}
\usepackage[colorlinks=true, pdfstartview=FitV, linkcolor=prdblue, citecolor= prdblue, urlcolor=prdblue]{hyperref}

\makeatletter

\setbox0\hbox{$\xdef\scriptratio{\strip@pt\dimexpr
    \numexpr(\sf@size*65536)/\f@size sp}$}

\newcommand{\scriptveryshortarrow}[1][3pt]{{%
    \hbox{\rule[\scriptratio\dimexpr\fontdimen22\textfont2-.2pt\relax]
               {\scriptratio\dimexpr#1\relax}{\scriptratio\dimexpr.4pt\relax}}%
   \mkern-4mu\hbox{\let\f@size\sf@size\usefont{U}{lasy}{m}{n}\symbol{41}}}}

\makeatother

\begin{document}

\begin{CJK*}{UTF8}{gbsn}

\title{$\,$\\[-6ex]\hspace*{\fill}{\normalsize{\sf\emph{Preprint no}.\ NJU-INP 092/24}}\\[1ex]%
Impressions of Parton Distribution Functions}

\date{2024 November 10}

\author{Yang Yu (俞杨)%
        $\,^{\href{https://orcid.org/0009-0008-8011-3430}{\textcolor[rgb]{0.00,1.00,0.00}{\sf ID}}}$}
\affiliation{School of Physics, \href{https://ror.org/01rxvg760}{Nanjing University}, Nanjing, Jiangsu 210093, China}
\affiliation{Institute for Nonperturbative Physics, \href{https://ror.org/01rxvg760}{Nanjing University}, Nanjing, Jiangsu 210093, China}


\author{Craig D. Roberts%
       $^{\href{https://orcid.org/0000-0002-2937-1361}{\textcolor[rgb]{0.00,1.00,0.00}{\sf ID}}}$}
\affiliation{School of Physics, \href{https://ror.org/01rxvg760}{Nanjing University}, Nanjing, Jiangsu 210093, China}
\affiliation{Institute for Nonperturbative Physics, \href{https://ror.org/01rxvg760}{Nanjing University}, Nanjing, Jiangsu 210093, China}

\begin{abstract}
\vspace*{-3ex}

\centerline{\small 
\href{mailto:cdroberts@nju.edu.cn}{cdroberts@nju.edu.cn} (CDR).}

\smallskip

Parton distribution functions (DFs) have long been recognised as key measures of hadron structure. Today, theoretical prediction of such DFs is becoming possible using diverse methods for continuum and lattice analyses of strong interaction (QCD) matrix elements.  Recent developments include a demonstration that continuum and lattice analyses yield mutually consistent results for all pion DFs, with behaviour on the far valence domain of light-front momentum fraction that matches QCD expectations. Theory is also delivering an understanding of the distributions of proton mass and spin amongst its constituents, which varies, of course, with the resolving scale of the measuring probe.  Aspects of the pion DF and proton spin developments are sketched herein, along with some novel perspectives on gluon and quark orbital angular momentum.\\[1ex]
\end{abstract}
%

\maketitle

\end{CJK*}


\noindent\emph{1.$\;$Introduction} ---
Quark partons were discovered almost sixty years ago in a series of experiments at the Stanford Linear Accelerator Center (SLAC), which used deep inelastic electron scattering to probe the structure of the nucleon \cite{Taylor:1991ew, Kendall:1991np, Friedman:1991nq, Friedman:1991ip}.  Theoretical developments following these experiments established that the measured cross-sections may typically be described by a convolution integral that is composed of a so-called hard scattering kernel, $K_{\rm hs}^{\rm ernel}$, which can, in principle, be calculated using perturbation theory within quantum chromodynamics (QCD -- the strong interaction piece of the Standard Model) and number density distributions, \emph{viz}.\ parton distribution functions (DFs).  We illustrate this here:
\begin{align}
\sigma(x,\zeta) & \sim \int_x^1\frac{dy}{y} \sum_{\rm partons, {\mathpzc p}} K_{\rm hs}^{\rm ernel}(y)\; {\mathpzc p}(x/y,\zeta)  \nonumber \\
& \quad + {\rm corrections}\,, \label{sigmadis}
\end{align}
where
${\mathpzc p}(x,\zeta)$ are the parton DFs and the ``corrections'' are small when the resolving energy scale satisfies $\zeta^2 \gg m_N^2$, with $m_N$ being the nucleon mass.

To be relevant to DF extraction, the kinematics of the deep inelastic scattering (DIS) process should be arranged to probe the Bjorken limit \cite{Bjorken:1968dy}: the virtual photon momentum ($q$) and the energy transfer to the nucleon target (whose momentum is $p$) must both be large ($m_N^2/q^2 \simeq 0$, $m_N^2/p\cdot q \simeq 0$); and yet the ratio $x = 2 p\cdot q/q^2$ should remain a fixed, finite number.   This variable is called ``Bjorken $x$''.

The key feature for a theoretical analysis of DIS (and kindred reactions) is that $x$ can directly be linked with the light-front fraction of the target's total momentum being carried by the parton which interacts with the energetic probe \cite[Ch.\,4.2]{Ellis:1991qj}.
This entails that ${\mathpzc p}(x,\zeta)$ is only nonzero on $x\in [0,1]$.

To understand these remarks, it is necessary to recall that one may choose to quantise any field theory using light-front coordinates \cite{Brodsky:1997de}.
(Light-front -- LF -- quantisation was introduced in Ref.\,\cite{Dirac:1949cp}.)
Further, that a principal merit of such an approach is that the derived light-front wave functions (LFWFs) possess a probability interpretation, much like a Schr\"odinger wave function.
This is not generally true of wave functions computed in other methods for quantisation of relativistic field theories.

Taking the LF approach to QCD, then a given hadron's parton DFs can be obtained from its LFWF \cite{Brodsky:1989pv}.
This connection can also be expressed using the concept of generalised parton distributions (GPDs) \cite{Diehl:2003ny}.
Importantly, one may also calculate a hadron's LFWFs indirectly, \emph{e.g}., via careful and appropriate projection of its Poincar\'e-covariant Bethe-Salpeter wave function \cite{Chang:2013pq}; and much progress has been made in the past decade using this method.
As projections of hadron wave functions, DFs are plainly both nonperturbative quantities and independent of the process used to measure them.

Before discussing theoretical prediction of DFs, it is worth stressing that the ability to calculate DFs is a recent development.
Today, as has been the case since the SLAC experiments, it is still common to infer DFs from fits to a collection of data on DIS and kindred reactions \cite{Ethier:2020way, Barry:2021osv, NNPDF:2021njg, Kotz:2023pbu}.
The reliability of this phenomenological approach rests on the capability of perturbation theory to supply a hard scattering kernel, $K_{\rm hs}^{\rm ernel}$ in Eq.\,\eqref{sigmadis}, which incorporates all important pieces.
Critically, however, without a solution of QCD, one does not know just what the crucial pieces are.
So, given that any phenomenological result for a set of DFs depends on the form used for $K_{\rm hs}^{\rm ernel}$, one cannot judge whether the fitted DF outcomes are an expression of QCD or, instead, merely a product of the truncated version of that theory which is represented by the hard scattering kernel employed.
Comparisons between the fits and theory DF predictions are therefore essential.
Conflicts may exist -- see, \emph{e.g}., Ref.\,\cite{Cui:2021mom}; and one can argue that they point to omissions in $K_{\rm hs}^{\rm ernel}$, which must be identified and remedied before existing and future data can be used to validate QCD.
This is becoming increasingly important as science enters an era of high-luminosity, high-energy facilities \cite{BESIII:2020nme, Anderle:2021wcy, Arrington:2021biu, Quintans:2022utc}.

\smallskip

\noindent\emph{2.$\;$Nambu-Goldstone Bosons} --- %
QCD is an extraordinary theory, with many fascinating features.  Its complexity has driven experiment and theory for fifty years \cite{Marciano:1979wa}; and significant progress has been made.  One basic fact is paramount; namely, the degrees of freedom used to write the QCD Lagrangian density -- bare/undressed gluon and quark partons -- are not the objects that reach detectors.  Gluon and quark confinement, a theoretically nebulous concept \cite[Sec.\,5]{Ding:2022ows}, entails that only QCD-charge (colour) neutral bound states can be measured.

Given confinement, then mesons, being systems comprised of just two valence degrees-of-freedom, would seem to be the simplest QCD objects to consider.  The lightest mesons are pions ($\pi$) and kaons ($K$), which are the ground states on the trajectories of mesons that can be built from either ($\pi$) one light quark ($u$, $d$) and one light antiquark or ($K$) a light quark/antiquark and a strange antiquark/quark ($\bar s$, $s$).  At issue here, however, is that, regarding typical QCD mass-scales, $\pi$ and $K$ are remarkably light; and this because they are Nature's most fundamental Nambu-Goldstone (NG) bosons \cite{Horn:2016rip}.

The dichotomous character of these systems, \emph{i.e}., their dual status as both bound-states and NG bosons, raises studies that can reveal $\pi$ and $K$ internal structure to the highest levels of importance.  Consequently, a raft of experiments is underway or anticipated in the foreseeable future \cite{BESIII:2020nme, Anderle:2021wcy, Arrington:2021biu, Quintans:2022utc}, with the goal of delivering data that can ultimately be used to chart the distributions of gluons and quarks there within.  A survey of experiment and theory is provided elsewhere \cite{Roberts:2021nhw}.  Herein, therefore, we sketch only a few more recent developments.  Complementary analyses and perspectives may be found elsewhere -- see, \emph{e.g}., Refs.\,\cite{Lan:2019rba, Han:2020vjp, Kock:2020frx, dePaula:2022pcb, Chang:2022pcb, Pasquini:2023aaf, Chang:2024rbs, Han:2024yzj}.

The ability of theory to deliver predictions for $\pi$ and $K$ DFs has grown substantially in the past decade.
This is partly because methods have been proposed and are being developed that may see the numerical simulation of lattice-regularised QCD (lQCD) begin to deliver realistic results for the $x$-dependence of DFs \cite{Ji:2013dva, Radyushkin:2017cyf, Ma:2017pxb}.  In that connection, however, issues must yet be resolved, such as ensuring the correct domain of support, $x\in[0,1]$, and/or solving or skirting the notorious ``inverse problem'' in physics, \emph{viz}.\ as in phenomenology, given $\sigma(x,\zeta)$, $K_{\rm hs}^{\rm ernel}$, how does one objectively extract an unknown function ${\mathpzc p}(x,\zeta)$ from an expression like Eq.\,\eqref{sigmadis}.

Consequently, pursuing a longstanding effort, lQCD continues to focus on the problem of computing DF Mellin moments:
\begin{equation}
{\mathpzc M}_{\mathpzc p}^m(\zeta) =\langle x^m \rangle_{\mathpzc p}^\zeta = \int_0^1 dx\,x^m  {\mathpzc p}(x;\zeta).
\end{equation}
Although more straightforward, there are nevertheless also challenges with such computations \cite{Holt:2010vj}.  For instance, the lQCD reach in $m$ is limited by achievable statistical precision and, at a more basic level, by the breaking of O$(4)$ symmetry -- the Euclidean version of Poincar\'e invariance -- introduced by lattice discretisation.
(The need for a Euclidean metric formulation of QCD is discussed elsewhere \cite[Sec.\,1]{Ding:2022ows}.)
Using local operators, lattice-spacing power-divergences in the mixing with lower-dimensional operators restricts access to $m\leq 3$ \cite{Alexandrou:2021mmi}.  Higher moments can be reached using nonlocal operators, or hybrid methods that exploit features of the frameworks introduced to make DF $x$-dependence available \cite{Joo:2019bzr, Sufian:2019bol}.  However, in some such schemes, only even moments of valence quark DFs are accessible \cite{Gao:2022iex}.  This is a serious drawback because the $m=1$ moment contains much important information, \emph{e.g}., it is the light-front momentum fraction carried by the parton, a key quantity for experiment, phenomenology, and theory.

Schwinger functions, which may also be called Euclidean space Green functions, lie at the heart of lQCD studies. There is also a longstanding effort to calculate these quantities in continuum QCD -- see Ref.\,\cite{Roberts:1994dr} and citations thereof; and significant advances have been made in the use of continuum Schwinger function methods (CSMs) during the past decade.  Notably, where reasonable comparisons are possible, contemporary CSM predictions and lQCD results are mutually consistent -- see, \emph{e.g}., Refs.\,\cite{Roberts:2021nhw, Binosi:2022djx, Ding:2022ows, Ferreira:2023fva, Raya:2024ejx, Chen:2021guo, Chang:2021utv, Lu:2023yna, Yu:2024qsd, Chen:2023zhh, Alexandrou:2024zvn}.

One illustration must suffice herein.  In order to draw it, some background to the all-orders (AO) evolution scheme for parton DFs will be helpful.  Its development began with the study of pion DFs in Ref.\,\cite{Ding:2019lwe} and a compact elucidation is presented in Ref.\,\cite{Yin:2023dbw}.  The basic point is that DFs depend on the probe resolving scale, $\zeta$.  The relevant value depends on the kinematics available to a given experiment.  Suppose that two experiments are performed, with values $\zeta$, $\zeta^\prime$, and these experiments enable extraction of ${\mathpzc p}(x;\zeta)$, ${\mathpzc p}(x;\zeta^\prime)$.  How can one tell whether the results are consistent, \emph{viz}.\ is there a mathematically precise connection between these two functions?  The answer is yes and the mathematical link is provided by the DGLAP evolution equations \cite{Dokshitzer:1977sg, Gribov:1971zn, Lipatov:1974qm, Altarelli:1977zs}.  Derived using perturbation theory in QCD, these equations map ${\mathpzc p}(x;\zeta) \leftrightarrow {\mathpzc p}(x;\zeta^\prime)$ on any domain for which perturbation theory is valid, \emph{i.e}., typically $\zeta^2, \zeta^{\prime 2}> m_N^2$.

The basic tenets of the AO scheme are simple.
(\emph{a}) There is an effective charge, $\alpha_{1\ell}(k^2)$, in the sense of Refs.\,\cite{Grunberg:1980ja, Grunberg:1982fw} and reviewed in Ref.\,\cite{Deur:2023dzc}, that, when used to integrate the leading-order perturbative DGLAP equations, defines an evolution scheme for \emph{every} parton distribution function (DF) that is all-orders exact.
The pointwise form of $\alpha_{1\ell}(k^2)$ is largely irrelevant.  Nevertheless, the process-independent strong running coupling defined and computed in Refs.\ \cite{Binosi:2016nme, Cui:2019dwv} has all requisite properties.
(\emph{b}) There is a scale, $\zeta_{\cal H}<m_N$, at which all properties of a given hadron are carried by its valence degrees-of-freedom.  At this scale, DFs associated with glue and sea quarks are zero.
The AO approach extends DGLAP evolution onto QCD's nonperturbative domain.

The AO scheme was recently used to answer some important theoretical questions relating to lQCD calculations of moments of the pion valence quark DF, ${\mathpzc u}^{\pi}_{\rm V}(x;\zeta)$.
(\emph{i}) Suppose one has collections of moments obtained from distinct lattice setups, at different resolving scales, and using dissimilar algorithms, then are they mutually consistent?
(\emph{ii}) If they are, is it possible to obtain a robust reconstruction of the DF, with reliable uncertainties,  from the available lQCD-determined Mellin moments?
In Ref.\,\cite{Lu:2023yna}, an approach to answering these questions was exemplified using the pion valence quark DF moments reported in Refs.\,\cite{Alexandrou:2021mmi, Joo:2019bzr, Sufian:2019bol, Gao:2022iex}, which are listed in Table~\ref{latticemoments}.


\begin{table}[t!]
\caption{
\label{latticemoments}
Lattice QCD results for Mellin moments of the pion valence-quark DF at
$\zeta=\zeta_2=2\,$GeV \cite{Joo:2019bzr, Gao:2022iex}
and
$\zeta_5=5.2\,$GeV \cite{Alexandrou:2021mmi, Sufian:2019bol}.
%
As discussed in connection with Eq.\,\eqref{EqGMin}, the column labelled ``G Eq.\,\eqref{EqGMin}'' provides the $\chi^2$ odd-moment completion of the Ref.\,\cite{Gao:2022iex} even moments.
}
\begin{tabular*}
{\hsize}
{
l@{\extracolsep{0ptplus1fil}}
|l@{\extracolsep{0ptplus1fil}}
l@{\extracolsep{0ptplus1fil}}
l@{\extracolsep{0ptplus1fil}}
l@{\extracolsep{0ptplus1fil}}
l@{\extracolsep{0ptplus1fil}}
l@{\extracolsep{0ptplus1fil}}}\hline\hline
$n\ $ & \cite[J]{Joo:2019bzr} & \cite[G]{Gao:2022iex} & G Eq.\,\eqref{EqGMin} &
\cite[A]{Alexandrou:2021mmi} & \cite[S]{Sufian:2019bol} \\\hline
$1\ $ & $0.254(03)\ $ & & $0.271\ $ & $0.23(1)\ $ & $0.18(3)\ $\\
$2\ $ & $0.094(12)\ $ & $0.1104(73)\ $ & &  $0.087(05)\ $ & $0.064(10)\ $\\
$3\ $ & $0.057(04)\ $ & & $0.054(8)\ $ & $0.041(04)\ $ & $0.030(05)\ $\\
$4\ $ & & $0.0388(46)\ $ &  & $0.023(05)\ $ & \\
$5\ $ & &  & $0.037(24)\ $ &  $0.014(04)\ $ & \\
$6\ $ & & $0.0118(48)\ $ & &   $0.009(03)\ $ & \\\hline\hline
%
%
%
%
%
%
%
\end{tabular*}
\end{table}

The moments in Ref.\,\cite[G]{Gao:2022iex} represent a novel case because the method used therein only delivers even moments, with good signals for $m=2, 4, 6$ -- see Table~\ref{latticemoments}.  To use the AO evolution scheme, the $m=1$ moment is required.  This was obtained in Ref.\,\cite{Lu:2023yna} using a $\chi^2$ minimisation procedure made with reference between the even moments in Refs.\,\cite{Alexandrou:2021mmi, Joo:2019bzr, Sufian:2019bol, Gao:2022iex} and using the value of the first moment in Ref.\,\cite[G]{Gao:2022iex} as the minimisation parameter.  This yielded
\begin{equation}
\label{EqGMin}
{\mathpzc M}_{\rm G}^1(\zeta_{\rm G}=\zeta_2) = 0.271\,,
\end{equation}
using which the AO scheme subsequently produces the $m=3, 5$ moments listed in Table~\ref{latticemoments}-col.\,3.  
It is worth stressing that the odd-moment completion of the even moments reported in Ref.\,\cite{Gao:2022iex} made neither assumptions about the form of the effective charge nor the value of the hadron scale.  It is therefore significant that using the PI charge elucidated in Ref.\,\cite{Cui:2019dwv}, $\hat \alpha(\zeta)$, along with the first moment from Ref.\,\cite{Alexandrou:2021mmi}, one finds agreement with Eq.\,\eqref{EqGMin}; and  repeating this procedure for all moments reported in Ref.\,\cite{Gao:2022iex}, one arrives at the following comparisons:
\begin{equation}
\begin{array}{c|ccc}
n & 2 & 4 & 6 \\\hline
{\mathpzc M}_{\rm G}^n(\zeta_{\rm G}) & 0.1104(73) & 0.0388(46)  & 0.0118(48) \\
{\mathpzc M}_{\rm A}^n(\zeta_{\rm G}) & 0.102(11)\phantom{1}  & 0.027(09)\phantom{1}  & 0.011(05)\phantom{1}
\end{array}\,.
\end{equation}
Evidently, the AO scheme reveals mutual consistency between these distinct sets of lQCD results.

The next question was: Now that the moments are known to be consistent, can they be used to reconstruct the pion valence quark DF?  Again, the answer is yes.  Further, the procedure is straightforward.  First, an important remark.  Namely, analyses of ${\mathpzc u}^{\pi}_{\rm V}(x;\zeta)$, which incorporate the behaviour of the pion wave function prescribed by QCD, predict \cite{Cui:2021mom}:
\begin{equation}
\label{pionDFpQCD}
{\mathpzc u}_{\rm V}^\pi(x;\zeta) \stackrel{x\simeq 1}{\sim} (1-x)^{\beta \,=\,2+\gamma(\zeta)}\,,
\end{equation}
where $\gamma(\zeta_{\cal H})=0$ and $\gamma(\zeta>\zeta_{\cal H}) \geq 0$ grows logarithmically with $\zeta$, expressing the physics of gluon radiation from the struck quark as encoded in DGLAP evolution.
The powers on glue and sea DFs are, respectively, one and two units greater \cite{Brodsky:1994kg, Yuan:2003fs, Holt:2010vj, Cui:2021mom, Cui:2022bxn, Lu:2022cjx}.


Informed by Eq.\,\eqref{pionDFpQCD} and using the AO scheme, Ref.\,\cite{Lu:2023yna} evolved all lQCD moments in Table~\ref{latticemoments} down to $\zeta_{\cal H}$.  Although the value of this scale is immaterial to the analysis, it is interesting to note that when working with the charge discussed in Refs.\,\cite{Cui:2019dwv, Deur:2023dzc, Brodsky:2024zev}, the value of the hadron scale is a prediction \cite{Cui:2021mom}:
$\zeta_{\cal H} = 0.331(2)\,{\rm GeV}$.  Using this same charge, Ref.\,\cite{Lu:2023yna} showed that the lQCD studies produce a consistent result: $\zeta_{\cal H} = 0.350(44)\,{\rm GeV}$.

Given Eq.\,\eqref{pionDFpQCD}, one may propose that hadron scale valence quark DFs can be described by the following one-parameter ($\rho$) form:
\begin{equation}
{\mathpzc u}^\pi(x;\zeta_{\cal H}) =
{\mathpzc n}_\pi \ln[ 1+x^2(1-x)^2/\rho^2]\,;
\label{CSMDF}
\end{equation}
and test the hypothesis using a $\chi^2$ procedure.
Then using all the $m\geq 2$ moments in Table~\ref{latticemoments}, one finds $\rho=0.061$ returns  $\chi^2/$degree-of-freedom$\,4.8/13=0.37$, \emph{viz}.\ a 98\% probability that the lQCD moments are described by the moments of Eq.\,\eqref{CSMDF}.

\begin{figure}[t]
\vspace*{0ex}

\leftline{\hspace*{0.5em}{{\textsf{A}}}}
\vspace*{-2ex}
\centerline{\includegraphics[width=0.86\columnwidth]{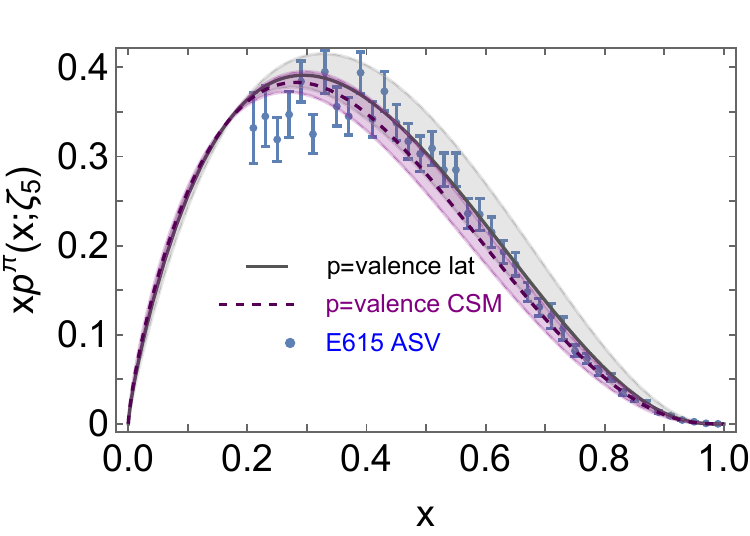}}
\vspace*{2ex}

\leftline{\hspace*{0.5em}{{\textsf{B}}}}
\vspace*{-2ex}
\centerline{\includegraphics[width=0.85\columnwidth]{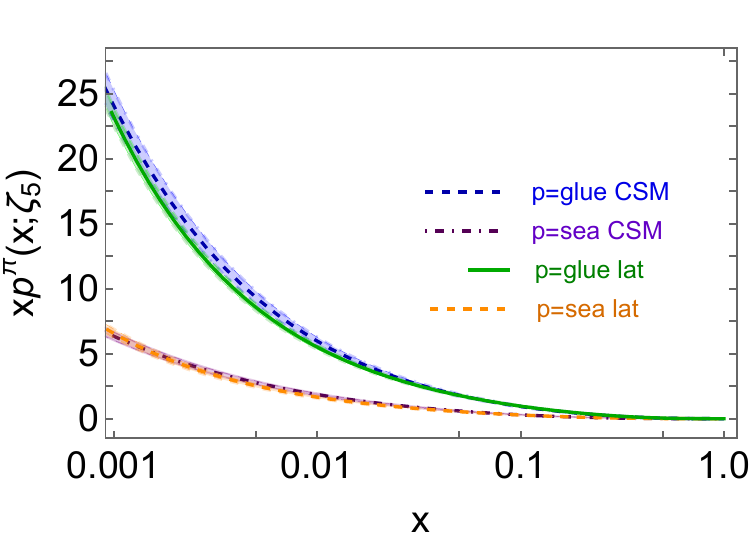}}

\caption{\label{piDFs}
{\sf Panel A}.
Lattice-QCD result (grey) for the pion valence quark DF at $\zeta=\zeta_5=5.2\,$GeV, reconstructed from the moments in Table~\ref{latticemoments} and evolved from $\zeta_{\cal H}$ using the AO scheme.
It is compared with the CSM prediction \cite{Cui:2020tdf} (purple) and the analysis of E615 Drell-Yan data described in Ref.\,\cite{Aicher:2010cb}.
{\sf Panel B}.
Glue and sea DFs at $\zeta_5$: lQCD compared with CSM predictions \cite{Cui:2020tdf}.
Like-coloured bands bracketing CSM results reflect the impact of a 5\% uncertainty in $\zeta_{\cal H}$.
}
\end{figure}

With this level of confidence, one may use the AO scheme to generate all pion DFs (valence, glue, total four-flavour sea) at any scale $\zeta>\zeta_{\cal H}$ from the lQCD valence quark DF.  A useful new scale is $\zeta=\zeta_5 = 5.2\,$GeV, at which pion + proton Drell-Yan reactions \cite{Drell:1970wh} ($\pi p \to \mu^+ \mu^- X$, with $X$ a shower of undetected particles) have been measured and therefrom ${\mathpzc u}^{\pi}_{\rm V}(x;\zeta_5)$ inferred \cite[E615]{Conway:1989fs}.   Figure~\ref{piDFs}A displays the valence DF at this new scale, compared with the CSM prediction \cite{Cui:2020tdf}: plainly, continuum and lattice results agree within mutual uncertainty; and, moreover, both agree with the analysis of E615 data described in Ref.\,\cite{Aicher:2010cb}.  Glue and sea DFs are drawn in Fig.\,\ref{piDFs}B: again, continuum and lattice agree.

Referring to Eq.\,\eqref{pionDFpQCD} and related remarks, it is important to note that the lQCD studies collected in Table~\ref{latticemoments} produce a large-$x$ exponent $\beta(\zeta_5) = 2.55 (36)$.  Analyses of experiment cannot be very certain about this $x\simeq 1$ (extreme endpoint) exponent.  For empirical purposes, therefore, it is better to report an effective exponent, \emph{i.e}., an average value on the domain $x\in [0.85,1.0]$ \cite{Holt:2010vj}.  Working with the results in Fig.\,\ref{piDFs}A, one finds
\begin{equation}
  \beta_{\rm valence}^{\rm eff-lQCD}(\zeta_5) \stackrel{x \in [0.85,1]}{=} 2.36(31)\,.
\end{equation}
Restricting the domain to $x\in [0.9,1.0]$, this value rises 2\%.
Analogous $x\in [0.85,1.0]$ results for glue and sea are: $  \beta_{\rm glue}^{\rm eff-lQCD}(\zeta_5) = 3.56(23)$, $\beta_{\rm sea}^{\rm eff-lQCD}(\zeta_5) = 4.62(26)$.

The analysis in Ref.\,\cite{Lu:2023yna} unifies modern continuum and lattice results for pion DFs with each other and shows that they align with longstanding predictions for the power-law behaviour of these functions on the far valence domain.  Kindred studies of the kaon exist and extensions to kaon lQCD results are underway.  Crucially, ongoing and anticipated experiments in a new era of high-luminosity accelerators are capable of validating the predictions sketched in this section \cite{BESIII:2020nme, Anderle:2021wcy, Arrington:2021biu, Quintans:2022utc}.

\smallskip

\noindent\emph{3.$\;$Proton Spin} --- %
CSMs are today providing a single-framework unification of pion, kaon, and nucleon properties -- see, \emph{e.g}., Refs.\,\cite{Yao:2024drm, Yao:2024uej, Yao:2024ixu}; and the past few years have seen the analyses of meson DFs, sketched above, extended to all nucleon DFs \cite{Chang:2022jri, Lu:2022cjx, Cheng:2023kmt, Yu:2024qsd}.
In this \emph{impression}, it is worth highlighting these nucleon applications via a discussion of the ``proton spin crisis''.
More exhaustive commentaries on related topics may be found elsewhere -- see, \emph{e.g}., Refs.\,\cite{Aidala:2012mv, Deur:2018roz}

The proton is a composite system, with unit positive electric charge, built from three valence quarks ($u+u+d$).  Its state vector is labelled by three Poincar\'e invariant quantities: mass squared, $m_p^2$; total angular momentum squared (spin), $J^2=J(J+1)=3/4$; and parity $P=+1$.  This seems straightforward.  However, it was long ago found that much of the proton's spin is not lodged with the spin of its valence quarks \cite{EuropeanMuon:1987isl}.  Like pions and kaons, therefore, proton structure holds surprises.  Herein, we will provide some insights into the spin puzzle.

From the AO evolution standpoint, all proton spin is carried by dressed valence degrees-of-freedom at $\zeta_{\cal H}$.  This is highlighted, \emph{e.g}., by Ref.\,\cite[Fig.\,3]{Liu:2022nku}:
(\emph{i}) The proton's Poincar\'e covariant wave function is obtained by solving a quark + fully-interacting diquark Faddeev equation.  (The quark + diquark approach to proton structure is explained in Ref.\,\cite{Barabanov:2020jvn}.)
(\emph{ii}) The solution transforms as a $J^P=\tfrac{1}{2}^+$ system, in whose rest frame there is substantial quark + diquark orbital angular momentum.
Thus, dressed valence quarks produce all of $J=\tfrac{1}{2}$, but part of that is associated with quark + diquark orbital angular momentum.
In a Poincar\'e invariant theory, this is unavoidable: Poincar\'e covariance of hadron wave functions demands the presence of orbital angular momentum in every frame; so, even the wave function of the $J=0$ pion has components with nonzero angular momentum \cite{Bhagwat:2006xi}.
Such outcomes occur irrespective of the approach used to calculate the wave function -- see, \emph{e.g}., modern three valence body studies of the proton \cite{Qin:2018dqp}.

Nonetheless, given the theoretical foundations of Eq.\,\eqref{sigmadis}, DF interpretation of measurements is impossible at $\zeta_{\cal H}$.  One must therefore evolve the Ref.\,\cite[Fig.\,3]{Liu:2022nku} picture to a scale that can be linked with experiments.  A good choice is that associated with the proton spin experiments described in Refs.\,\cite[COMPASS]{COMPASS:2010hwr, COMPASS:2016jwv}, \emph{viz}.\ $\zeta=\zeta_{\rm C} = \surd 3\,$GeV.

Beginning at $\zeta_{\cal H}$, the quark + diquark approach to proton structure produces the following light-front decomposition of the proton spin into contributions from quark and gluon spin and orbital angular momenta (OAM) \cite{Jaffe:1989jz}:
{\allowdisplaybreaks
\begin{equation}
\frac{1}{2}  = \frac{1}{2}
\sum_{{\mathpzc q} = {\mathpzc u}, {\mathpzc d}} \langle \Delta {\mathpzc q_p} \rangle^{\zeta_{\cal H}}
+ \Delta G(\zeta_{\cal H})
+ \sum_{{\mathpzc q} = {\mathpzc u}, {\mathpzc d}} \ell_{\mathpzc q}(\zeta_{\cal H})
+ \ell_g(\zeta_{\cal H})\,,
\end{equation}
where \cite{Cheng:2023kmt, Yu:2024qsd}
\begin{subequations}
\label{spinbreak}
\begin{align}
& {\rm quark}{\rm ~helicity\!:} & \nonumber \\
& \qquad a_{\rm I} = \sum_{{\mathpzc q} = {\mathpzc u}, {\mathpzc d}} \langle \Delta {\mathpzc q_p} \rangle^{\zeta_{\cal H}}
 = 0.52(1)\, g_A = 0.65_{\pm 0.01}\,, \label{QH}\\
& {\rm glue~helicity\!:} \;  \Delta G(\zeta_{\cal H}) = 0\,, \\
& {\rm quark~OAM\!:} \; \sum_{{\mathpzc q} = {\mathpzc u}, {\mathpzc d}} \ell_{\mathpzc q}(\zeta_{\cal H}) = 0.175_{\mp 0.01} \,, \label{qOAMA}\\
& {\rm glue~OAM\!:} \; \ell_g(\zeta_{\cal H}) = 0\,.
\end{align}
\end{subequations}
}

Here it should be stressed that any decomposition of total angular momentum, $J$, into spin, $S$, plus orbital angular momentum, $L$, is frame and scale dependent: neither $S$ nor $L$ is Poincar\'e-invariant and, in quantum field theory, the degrees-of-freedom used to perform any such decomposition change with scale.  Hence, neither $S$ nor $L$ is truly observable.  One may make reference to them only after fixing the observer's frame -- a light-front perspective is adopted herein  -- and specifying the measuring scale, $\zeta$.

At $\zeta_{\cal H}$, the individual light-front flavour contributions, $\ell_{{\mathpzc u},\mathpzc{d}}$ in Eq.\,\eqref{spinbreak}, may be estimated using a Wand\-zura-Wilczek approximation \cite{Hatta:2012cs, Bhattacharya:2023hbq}:
\begin{equation}
\label{LWW}
\ell_{\mathpzc f}(x;\zeta_{\cal H}) =
x \int_x^1 dy \frac{1}{y^2}
\left[ y{\mathpzc f}(y;\zeta_{\cal H}) - \Delta{\mathpzc f}(y;\zeta_{\cal H})\right]\,,
\end{equation}
where ${\mathpzc f}$, $\Delta{\mathpzc f}$ are unpolarised, polarised $f$-flavour DFs.
In order to avoid relying on poorly constrained phenomenology, this expression omits a term associated with $E^q$, \emph{i.e}., the GPD connected with the proton Pauli form factor.

\begin{figure}[t]
\vspace*{0ex}

\leftline{\hspace*{0.5em}{{\textsf{A}}}}
\vspace*{-3ex}
\centerline{\includegraphics[width=0.86\columnwidth]{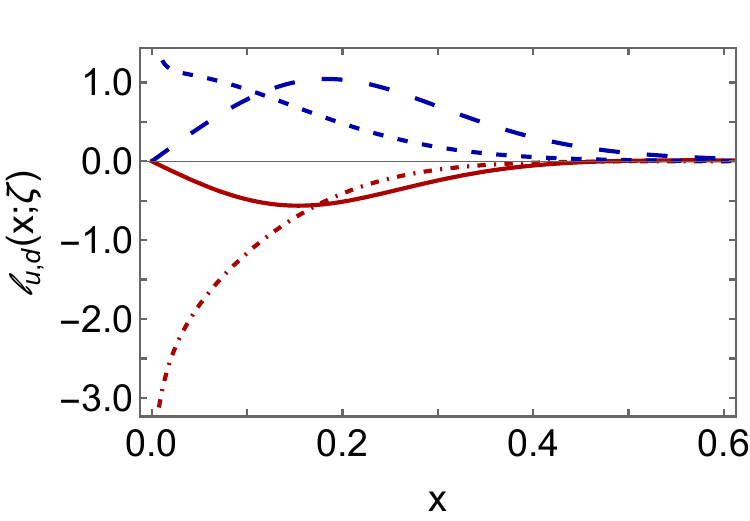}}
\vspace*{2ex}

\leftline{\hspace*{0.5em}{{\textsf{B}}}}
\vspace*{-3ex}
\centerline{\includegraphics[width=0.85\columnwidth]{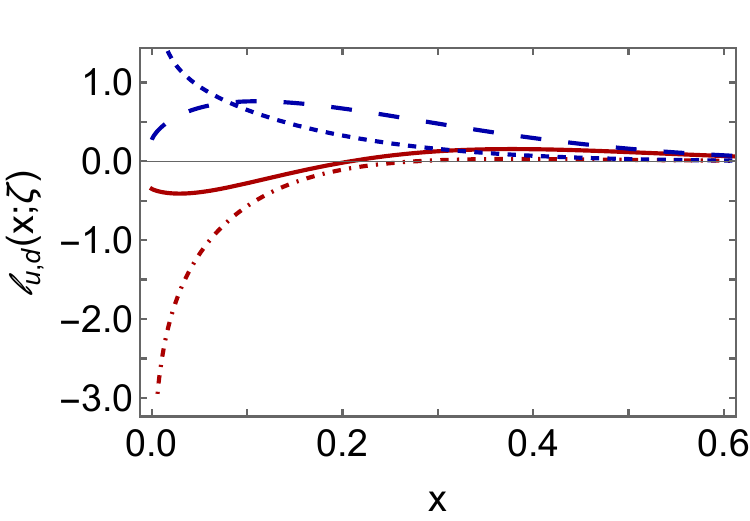}}

\caption{\label{quarkOAM}
{\sf Panel A}.
Light-quark orbital angular momentum DFs calculated using Eq.\,\eqref{LWW} and the DFs in Ref.\,\cite{Cheng:2023kmt}.
Legend.
$\zeta_{\cal H}$: $u$ quark -- solid red curve; and $d$ quark -- long-dashed blue curve.
$\zeta_{\rm C}$: $u$ quark -- dot-dashed red curve; and $d$ quark -- dashed blue curve.
{\sf Panel B}.  Same as {\sf A}, except that the contact interaction DFs in Ref.\,\cite{Yu:2024qsd} were used as input.
}
\end{figure}

Using Eq.\,\eqref{LWW} and the DFs in Ref.\,\cite{Cheng:2023kmt}, one obtains the following results for the zeroth Mellin moments:
\begin{equation}
\ell_u(\zeta_{\cal H}) = -0.134\,, \quad \ell_d(\zeta_{\cal H})=0.309\,,
\label{ellud}
\end{equation}
along with the profiles in Fig.\,\ref{quarkOAM}A.
In a test of Eq.\,\eqref{LWW}, the values in Eq.\,\eqref{ellud} sum to a result that precisely matches the prediction in Eq.\,\eqref{qOAMA}.  Crucially, this value was calculated without reference to Eq.\,\eqref{LWW}.  It is obtained simply because the Poincar\'e-covariant quark + diquark wave function transforms as a $J=1/2$ spinor and the same  wave function generates the axial charge sum in Eq.\,\eqref{QH}.
Regarding the profiles, both valence OAM DFs are small on $x\gtrsim 0.6$: the $d$ quark OAM is nonnegative, whereas $\ell_u(x;\zeta_{\cal H})$ is negative on $0<x<0.52$.

\begin{figure}[t]
\vspace*{0ex}

\leftline{\hspace*{0.5em}{{\textsf{A}}}}
\vspace*{-2ex}
\centerline{\includegraphics[width=0.86\columnwidth]{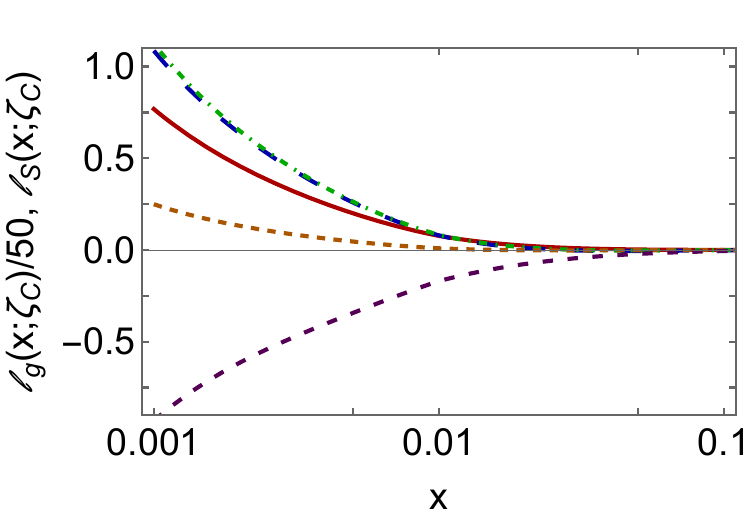}}
\vspace*{2ex}

\leftline{\hspace*{0.5em}{{\textsf{B}}}}
\vspace*{-2ex}
\centerline{\includegraphics[width=0.85\columnwidth]{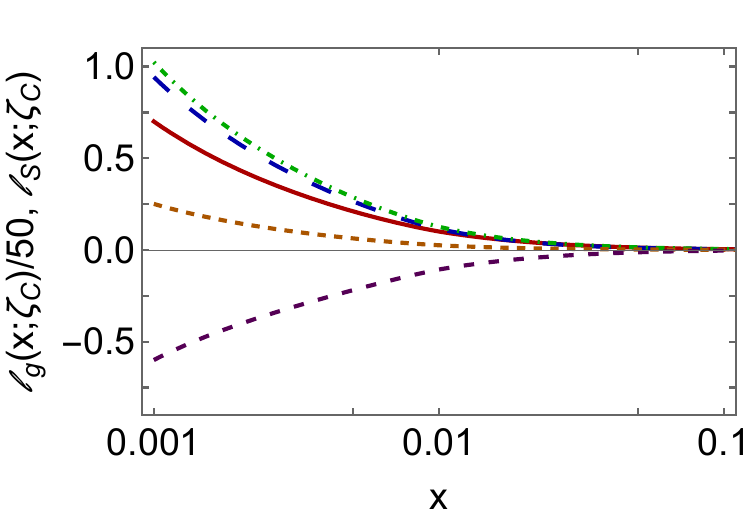}}

\caption{\label{quarkOAM2}
{\sf Panel A}.
Nonvalence quark and glue orbital angular momentum profiles, calculated using Eq.\,\eqref{LWW} with DFs generated via AO evolution to $\zeta_{\rm C}$ of the hadron-scale DFs in Ref.\,\cite{Cheng:2023kmt}.
Legend.
Solid red -- $2\ell_{\bar {\mathpzc u}}(x;\zeta_{\rm C})$;
long-dashed blue -- $2\ell_{\bar {\mathpzc d}}(x;\zeta_{\rm C})$;
dot-dashed green -- $\ell_{{\mathpzc s}+\bar {\mathpzc s}}(x;\zeta_{\rm C})$;
short-dashed orange -- $\ell_{{\mathpzc c}+\bar {\mathpzc c}}(x;\zeta_{\rm C})$;
and dashed purple -- $\ell_{\mathpzc g}(x;\zeta_{\rm C})/50$.
{\sf Panel B}.
As in {\sf A}, except that the contact interaction DFs in Ref.\,\cite{Yu:2024qsd} were used as input.
}
\end{figure}

Using AO evolution \cite{Yin:2023dbw}, implemented with straightforward modifications of Eqs.\,(32) in Ref.\,\cite{Deur:2018roz}, one finds (central values):
\begin{equation}
\ell_q(\zeta_{\rm C}) = -0.027\,,\,
\Delta G(\zeta_{\rm C}) = 1.30\,,\,
\ell_g(\zeta_{\rm C}) =-1.09\,.
\label{SpinBudget}
\end{equation}
The quark helicity contribution in Eq.\,\eqref{QH} is invariant under evolution.

Whilst remaining a useful qualitative guide, the quantitative accuracy of Eq.\,\eqref{LWW} degrades as the glue contribution increases.  Nevertheless, the analysis in Ref.\,\cite{Yu:2024qsd} suggests that Eq.\,\eqref{LWW} is valid at the $\sim 80$\% level for $\zeta=\zeta_{\rm C}$.  At this scale, one obtains the evolved valence quark OAM DFs in Fig.\,\ref{quarkOAM}A.
Evidently, under evolution, using Eq.\,\eqref{LWW}:
quark OAM is shifted to lower $x$;
both $u$ and $d$ quark OAM grow in magnitude;
the $d$ contribution becomes increasingly large and positive on its shrinking domain of material support;
and the $u$ profile grows even more rapidly negative.

It is worth highlighting that although valence quark OAM profiles are sensitive to details of proton structure, the impacts are typically quantitative, not qualitative.  This is elucidated by the curves in Fig.\,\ref{quarkOAM}B, which were obtained from DFs calculated using a symmetry-preserving contact-interaction study of proton structure  \cite{Yu:2024qsd}.


Herein, we also use Eq.\,\eqref{LWW} to provide estimates at $\zeta_{\rm C}$ for sea OAM DF profiles and the following analogue for the glue profile \cite{Hatta:2012cs}:
\begin{equation}
\label{LWWg}
\ell_{g}(x;\zeta_{\rm C}) =
x \int_x^1 dy \frac{1}{y^2}
\left[ \tfrac{1}{2} y{G}(y;\zeta_{\cal C}) - \Delta{G}(y;\zeta_{\cal C})\right]\,.
\end{equation}
This procedure is justified by noting, \emph{e.g}., that, obtained in this way, the predicted $u$, $d$, $s$, $c$ quark orbital angular momenta are consistent with the lQCD results in Ref.\,\cite{Alexandrou:2020sml} -- for details, see the discussion of Eq.\,(49) in Ref.\,\cite{Yu:2024qsd}; and the glue DF produced by Eq.\,\eqref{LWWg} yields $\ell_g(\zeta_{\rm C})$ in line with that reported in Eq.\,\eqref{SpinBudget}.

Figure~\ref{quarkOAM2} depicts sea and glue OAM DF profiles obtained using AO evolution and the OAM estimators in Eqs.\,\eqref{LWW}, \eqref{LWWg}.
Evidently, sea and glue OAM are concentrated at low $x$; and comparing the two panels, this feature is independent of whether realistic or contact interaction results are used for the hadron scale DFs.
Notably, each sea OAM DF is positive.
This is consistent with the lQCD results in Ref.\,\cite{Alexandrou:2020sml}, which is a valid comparison when Eqs.\,\eqref{LWW}, \eqref{LWWg} are good approximations \cite{Hatta:2012cs}.
On the other hand, the glue OAM DF is negative; and the magnitude of the glue OAM DF far exceeds that of the analogous sea DFs.  (This is also true of both the kindred unpolarised and helicity DFs.)
Before moving on, it should be stressed that the discussion of OAM has long been active and contentious -- see, e.g., Refs.\,\cite{Jaffe:1989jz, Ji:1996ek, Chen:2008ag} and citations thereof, and the CSM perspective presented here, drawn in part from Ref.\,\cite{Yu:2024qsd}, will be developed further in future.

These remarks highlight that it should never have expected that the nucleon spin is merely the sum of constituent/valence quark spins, which is the perspective that led to the ``proton spin crisis''.  Furthermore, it should now be plain that every proton spin decomposition depends on the practitioner's standpoint.  So, what should measurements of the proton spin yield?

As always, each measurement relates to the current used; and proton spin studies measure a conserved current that connects directly to the zeroth Mellin moment of the polarised proton structure function $g_1(x;\zeta)$.  This current is a sum over quark flavours; namely, a flavour singlet.  Consequently, it mixes with glue \cite{Altarelli:1988nr}.  Mapped into the notation used herein and employing AO evolution, the quantity measured in this way is the proton flavour-singlet axial charge:
{\allowdisplaybreaks
\begin{subequations}
\label{Eqa0E}
\begin{align}
a_{\rm I}^{\rm E}(\zeta) & = a_{\rm I} - n_f \frac{\hat \alpha(\zeta)}{2\pi} \int_0^1 dx\,\Delta G(x;\zeta) \\
& =: a_{\rm I} - n_f \frac{\hat \alpha(\zeta)}{2\pi} \Delta G(\zeta)\,,
\end{align}
\end{subequations}
where $n_f=4$ is the number of active quark flavours.
Crucially, CSMs predict that $\Delta G(x;\zeta) > 0$ on $x\in [0,1)$ and both terms on the right-hand side of Eq.\,\eqref{Eqa0E} are $\zeta$-independent under AO evolution; hence, the measured value of the in-proton quark helicity may receive a (material) reduction from the gluon helicity.
}

Using the predictions in Eqs.\,\eqref{QH}, \eqref{SpinBudget}, one finds
\begin{equation}
a_{\rm I}^{\rm E}(\zeta_{\rm C}) = 0.34(1).
\end{equation}
The measured value of the proton helicity is \cite[COMPASS]{COMPASS:2016jwv}: $0.32(7)$.
Herefrom, one sees that contemporary CSM analyses deliver a viable, parameter-free resolution of the ``proton spin crisis'' -- see, also, Ref.\,\cite[Fig.\,4]{Yin:2023dbw}.

It is now plain that, viewed from the light-front at resolving scales typical of modern measurements,
40\% of the proton spin is lodged with glue total angular momentum and 60\% with quarks (see also Ref.\,\cite{Yao:2024ixu}) although not all that quark contribution is apparent in the value of $a_{\rm I}^{\rm E}(\zeta_{\rm C})$.
The glue and quark fractions change with increasing $\zeta$, approaching the asymptotic ($m_N/\zeta \simeq 0$) values discussed in Refs.\,\cite{Ji:1995cu, Chen:2011gn}.
(Similar statements may be made about the light-front momentum fractions \cite{Lu:2022cjx}.)

\smallskip

\noindent\emph{4.$\;$Summary and Outlook} --- %
Experiments, underway and planned worldwide, will deliver data intended for use in the extraction of hadron parton distribution functions (DFs).
The focus is intense because knowledge of DFs can provide keen and unique insights into hadron structure.
For instance, theory progress in the past five years has delivered predictions for pion, kaon, and proton DF pointwise behaviour, which have revealed that pion DFs are the hardest (most dilated) in Nature \cite{Lu:2022cjx} as a consequence of strong interaction mechanisms that underlying the emergence of $\gtrsim 98$\% of visible mass in the Universe \cite{Roberts:2021nhw, Binosi:2022djx, Ding:2022ows, Ferreira:2023fva, Raya:2024ejx}.
Furthermore, studies of proton spin and how it is shared amongst the proton's constituents have indicated that roughly 40\% is carried by glue at resolving scales typical of modern measurements, with similar statements true of the proton mass \cite{Yao:2024ixu}.  Viewed from certain standpoints, quark and gluon orbital angular momenta (OAM) also carry material fractions of the proton's total spin, with OAM DF support concentrated at low values of parton light-front momentum fraction.

Precise data and reliable phenomenology are required to validate the growing array of theory DF predictions and derived insights.  The approaching era of high-luminosity, high-energy facilities will deliver the necessary data.  Increased feedback and synergy between phenomenology and theory, both continuum and lattice, is necessary to build a solid bridge between that data and quantum chromodynamics, the strong interaction piece of the Standard Model.

%
\medskip
\noindent\emph{Acknowledgments}.
%
%
Work supported by:
National Natural Science Foundation of China (grant no.\,12135007).
%
%


\end{document}